# Heliyon

# Feasibility study of antenna synthesis using hyper beamforming

Gebrehiwet Gebrekrstos Lema[*], Dawit Hadush Hailu

School of Electrical and Computer Engineering, Mekelle University, Mekelle, Ethiopia

[*] Corresponding author.

E-mail addresses: g.jcool.com@gmail.com, gebrehiwet.gebrekrstos@mu.edu.et (G.G. Lema).

## Abstract

Wireless communication requires an effective antenna synthesis that characterizes adequate infrastructures to provide the broader bandwidth and reduced interference. Antenna design with minimal signal degradation, optimal gain directive main beam to sustain minimal loss has been a hot issue among many communication engineers for several years. In this paper, the effects of eccentricity of the antenna, element-spacing, number of elliptical rings and number of elements are evaluated. For efficient antenna synthesis, deep Side Lobe Level (SLL) reduction and superior directivity are critical. We have also studied the significance of hyper beamforming in Elliptical Cylindrical Antenna Array (ECAA) in comparison to the geometric configuration of the antenna parameters (eccentricity of the antenna, element-spacing, number of elliptical rings and number of elements). The hyper beam exponent has resulted in flexible pattern synthesis while simultaneously reducing the side lobe of the proposed antenna array, thus decreasing the SLL and increasing directivity that are vital for wideband applications.

Keyword: Electrical engineering

## 1. Introduction

In recent years, the wireless communications seeking highly directive radiation characteristics as well as enormous reduction of the antenna side lobe is becoming crucial requirement of many researchers.

Many optimization algorithms replace the bulky or even impossible manual antenna designs. Among those algorithms the Self-adaptive Differential Evolution (SaDE), Genetic Algorithm (GA) and Particle Swarm Optimization (PSO) have shown great importance [1, 2, 3, 4]. Reduces the SLL keeping beam width unchanged. The SLL reduction in linear antenna arrays [5] opened the path to: amplitude only control, phase only control, both amplitude and phase control for effective design. To obtain a desired SLL reduction, calculating the antenna parameters and defining antenna patterns have continued a hot issue of many antenna array designers. An effective, robust, and simple global optimization algorithm called differential evolution comes in effect [6]. It has proven much better performance than many other evolutional algorithms in terms of convergence speed, robustness over several target functions and factual world troubles [7]. SaDE on the other hand, simultaneously adopts more than one mutation design [8]. An Improved Particle Swarm Optimization (IPSO) [9] is used to find optimal sets of non-uniformly excited hyper beamforming of linear antenna arrays. Similarly, an elliptical antenna array pattern is performed using SaDE [10] and it has resulted in a good side lobe reduction. But for wideband applications, these SLL reduction and directivity are insufficient.

In 2013, Amir Samanzare [11] has performed elliptical antenna array pattern synthesis. Here, though the GA method was used to adjust elements phases in the elliptical antenna array to obtain reduced SLL having good directivity, the SLL reduction is less than 10 dB over the non-optimized one. This indicates that the GA-based directivity improvement and SLL reduction are not substantial enough. Elliptical array has one extra parameter called eccentricity as compared to other geometry which helps to reduce SLL [12]. However, it is seen that with the variation of the eccentricity, the results do not show significant directivity increase or SLL reduction.

To achieve more determining geometric parameters, the elliptical cylindrical antenna array is formed by hybridizing linear antenna and elliptical antenna arrays [13]. In 2014, Rajesh Bera *et.al* [14] has designed the ECAA and He is able to reduce the SLL from -13dB to -31.72 dB using PSO. However, this SLL reduction is insufficient to be effectively utilized in high signal quality seeking applications like in radar systems, satellite communication and deep space explorations. The limitations of [14] are: The SLL reduction is not enough for long distance requiring services, the directivity is insufficient for point-to-point and satellite communications and it didn't plant any mechanism of pattern flexibility.

To overcome the limitations defined above, the ECAA is enhanced in the geographical configuration and we have proposed a different designing technique. The geometrical configuration includes eccentricity of the antenna, element-spacing, number of elliptical rings and number of elements and we have used a hyper beamforming unlike PSO. The hyper beam involvement increases the received signal available at the inputs of the receiver and increases the effective radiated power [15]. Finally, the directivity of the beam antenna allows the operator to null interfering stations.

The proposed geometrical configuration enables steering electronically, which allows scanning very rapidly across a wide angle of azimuth and elevation which therefore provides immediate observation over wide coverage area [16, 17]. The hyper beamforming-based SLL reduction, FNBW reduction and radiation pattern flexibility are significant compared to the different techniques mentioned in the literatures including GA, PSO and SaDE optimized hyper beam. However, it is possible to enhance the antenna design further if the hyper beamforming is hybridized with other optimization algorithms. Even though wider bandwidth is possible in guided media, wireless communication is much better in terms of cost and simplicity [18]. However, to achieve these advantages, reducing the antenna side lobe is critical as this increases the intensity of the radiation pattern, and hence decrease interference which in turn creates less packet loss wireless communication. Using symmetric array distribution, it is good that the scanning range is improved on thinned antenna array [19] but the degree of SLL reduction is not enough for wideband applications.

In this study, in addition to the capability of the antenna scanning over a wide coverage, the flexibility of the radiation pattern is used as a best reason for wideband applications, as described in Section 2. The antenna array excitation elements are uniform with normalized plots, all set to 1.

In this paper, the study was performed using hyper beamforming in comparison to the antenna configuration parameters effects for the enhancements of antenna radiation characteristics. The antenna configuration parameters are vertical spacing, ellipticity of the proposed configuration and number of array elements. Furthermore, this feasibility study was conducted with only the beamforming technique unlike the current literatures in which they use either combination of algorithms or optimization algorithms to design the proposed antenna. The previous works [16], and [17], didn't also study the antenna configuration parameters effects (vertical spacing, ellipticity and number of array elements) on the radiation pattern enhancement of the proposed antenna. The closest paper to this work is [13]. In [13], the effect of ellipse eccentricity, inter-element spacing and number of elements are evaluated against different current distributions. However, since that paper didn't apply the concept of beamforming, the SLL reduction was trivial.

## 2. Methods

In this paper, the methods used to evaluate the antenna performance are characterized by the evidences: the eccentricity of the antenna, the element-spacing, the number of elliptical rings and the number of elements for a high-performance

 

communication system, side lobe reduction and improvement in detecting the target is the goal of successful antenna designers. The classic way of beamforming increases the antenna array elements, the size and cost of the antenna becomes complex. Hence, a new technique called hyper beam is introduced in this paper that contributes to reduced SLL and increased directivity that are the hallmark for increasing the Signal-to-Noise ratio (SNR) which in turn reduces the interference. The hyper beamforming processing improves significantly the gain of the wireless link over a conventional technology, thereby increasing range, rate, and penetration capabilities of the signal.

Hyper beam is derived from sum and difference beam patterns each raised to the power of the hyper beam exponent parameter. Hyper beam method of beamforming generates a narrow beam as compared to conventional beam and results in improved performance of SLL and directivity that depend on the variation of exponent parameter value.

The hyper beam technique yields the narrow beams as a result of the spatial signal processing as shown in Fig. 1(a). The objective of the hyper beamforming to increase the directivity and suppress the SLL. This also leads to the suppression of grating lobe and reduction of received noise level.

In general, the detrimental motivations to use hyper beam technique are: (i) Simultaneous beam width and side lobe suppression, (ii) Grating lobe suppression for certain array configurations, (iii) Allowing pattern flexibility through hyper beam exponent.

## 3. Model

An ECAA is a hybrid of linear antenna array and elliptical antenna array which is composed of a number of parameters working together to create the required radiation pattern. The antenna array is designed to generate a pattern that is very low side lobe levels keeping the first null beam width as small as possible.

In the hyper beamforming, the phases of the antenna elements are organized in the manner that they will add up in the direction of interest while destructively interfering in the non-interested direction. Hyper beamforming of the antenna, considering the inter-element spacing of $0.5\lambda$, the sum beam can be created by summation of the absolute values of complex left and right half beams, as shown in Fig. 1 (b).

On the other hand, the difference beam is the magnitude of the absolute of the difference of complex right half beam and left half beam signals. Furthermore, the difference beam has a minimum in the direction of the sum beam at zero degree as shown in Fig. 1 (c). Thus, the resultant hyper beam will be obtained by subtracting

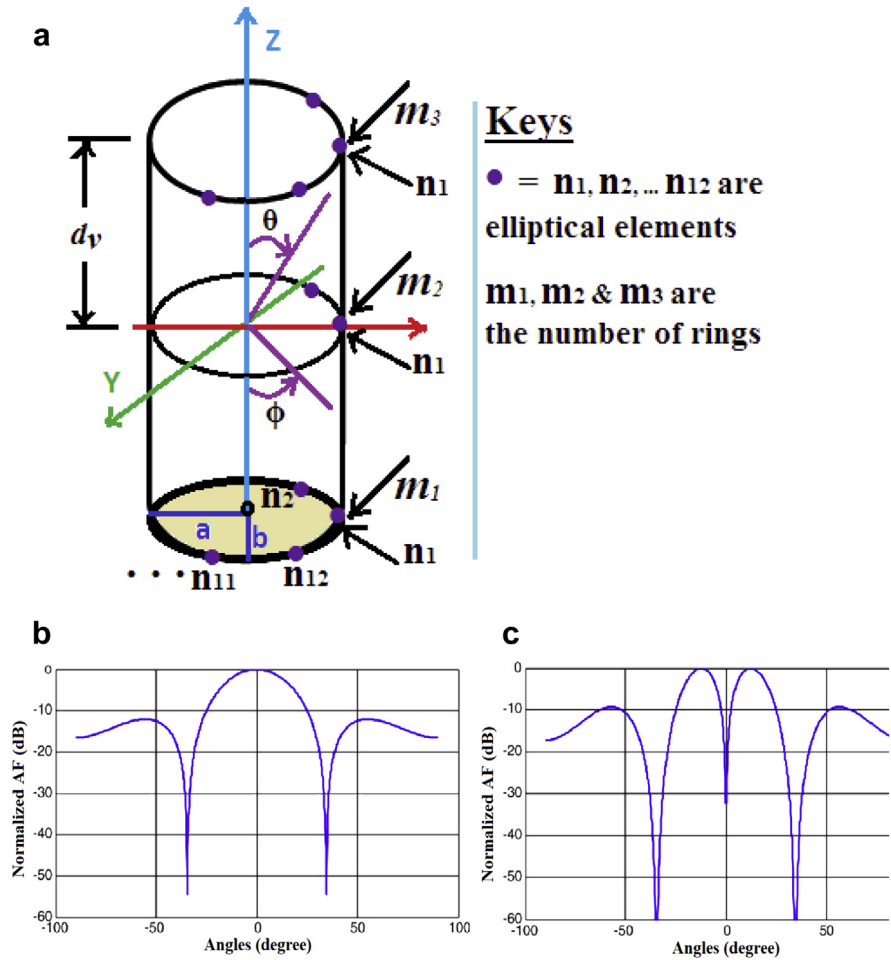

**Fig. 1.** (a): Geometrical configuration ECAA. (b): Hyper beam exponent of 1 for the sum of the left- and right-hand side beams. (c): the difference of the left- and right-hand side beams.

of the sum and the difference of the tow beams, each raised to the power of the exponent r.

More briefly, the sum beam can also be created by complex summation of the left and right half of the beams. The beam magnitude is equal to the left and right half beams, and it is also the equal to the sum of the magnitudes of both half beams if we consider the normalized Array Factor (AF) versus angles plot. The difference beam is magnitude of the difference of the right half beam signal subtracted from the left half beam signal by considering also the phases of the signals. It is obvious that the values of the difference beam at each given direction are always lower or equal to those of the half beam. Note that the difference beam has a minimum in the direction of the maximum sum beam. With this initial information, it is straight forward the idea of subtracting the magnitude of the difference pattern from the half beam pattern.

This technique is mathematically formulated as the difference and the sum of the radiation pattern(R) of the two half beams are as follows

D $(\theta,\phi)_{difference} = |R_{Left}-R_{Right}|$ and $S_{sum}(\theta,\phi) = |R_{Left}|+|R_{Right}|$ respectively. Then, the hyper beam equation with hyper beam exponent (r) is: $R_{hyper} = \{(|R_{Left}|+|R_{Right}|)^r - (|R_{Left}-R_{Right}|)^r\}^{1/r}$. The sum and difference patterns of the left beam and right beam are derived from the array factor of the proposed antenna array described as (1):

$$AF_{ECAA}(\theta,\phi) = \sum_{m=1}^{M}\sum_{n=1}^{N} I_{mn} e^{j(k\sin\theta(a\cos\phi\times\cos\phi_n + b\sin\phi\times\sin\phi_n))} \quad (1)$$

Where

$$R_{Right} = \sum_{m=1}^{M}\sum_{n=\frac{N}{2}+1}^{N} I_{mn} e^{j(k\times\sin\theta(a\cos\phi\times\cos\phi_n + b\sin\phi\times\sin\phi_n))} \quad (2)$$

$$R_{Left} = \sum_{m=1}^{M}\sum_{n=1}^{\frac{N}{2}} A_{mn} e^{j(k\sin\theta(a\cos\phi\times\cos\phi_n + b\sin\phi\times\sin\phi_n))} \quad (3)$$

$$A_{mn} = A_m \times e^{j((m-1)(k\times dv\times Cos\theta + p_m))} A_n \times e^{jp_n}$$

$A_n$ is the excitation amplitude of the $n^{th}$ element of the elliptical component, $A_m$ is the excitation currents of the $m^{th}$ element of the linear array. $P_m = k.dv.\cos(\theta_0)$, excitation phase shift in the linear component of the array and $P_n = k.\sin(\theta_0)(a.\cos(\phi_n).\cos(\phi_0) + b.\sin(\phi_n).\sin(\phi_0))$ is excitation phase shift in the elliptical component of the array. Where $\theta$ is the angle of radiation of electromagnetic plane wave and $d_v$ is the vertical inter-element spacing. The terms $\theta_0$ and $\phi_0$ are the angles in the main beam direction. $\phi_n = 2\pi(n-1)/N$ is the angular position of the $n^{th}$ element of the elliptical component in the xy-plane. $k = 2\pi/\lambda$ is the wave number. M is the number of linear arrays extended parallel to the z-axis, N is number of elements present on $m^{th}$ array and $A_{mn}$ is the amplitude excitation of the $n^{th}$ and $m^{th}$ elements.

## 4. Design

The simulation of this paper is based on the parameters given in Table 1. The simulation parameters are chosen when the study has shown better antenna radiation characteristics using the pseudo code given in Section 5. The radiation characteristics of some of the parameters is also discussed in Section 5.

The sum of the beam is created by the summation of the absolute of the complex values of the left and right half beams. Similarly, the difference of the beam is determined as the absolute magnitude of the difference of complex values of right half beam and the left half beam signals. Finally, the overall resulting hyper beam is obtained by subtraction of sum and difference beams, each raised to the power of the exponent **r**. The value of **r** is used to provide radiation flexibility.

**Table 1.** Plot parameters.

| Parameter | Symbol | Parameter value |
|---|---|---|
| Exponent of the hyper beam | r | Variable |
| Eccentricity of the ellipse | e | 0.5 (varying) |
| Elliptical longer axis | a | 1.15 (variable) |
| Elliptical shorter axis | b | 0.99 |
| Frequency | f | 305 MHz |
| Rings spacing | $d_v$ | 0.5λ |
| linear array m$^{th}$ element | M | 3 (varying) |
| Elliptical array $n^{th}$ element | N | 12 (varying) |
| Amplitude of the m$^{th}$ element | **A$_m$** | Fixed |
| Amplitude of the n$^{th}$ element | **A$_n$** | Fixed |

## 5. Results and discussions

In this paper, the suitability of the antenna for wideband application is studied using two groups of parameters. The first discusses about the effects of geometrical configuration of the antenna. These parameters are given in A through C: (A) the vertical spacing, (B), number of elliptical rings, (C) ellipse major-axis (i.e. the ellipse eccentricity), (D) number of elements. The second group investigates about the spatial signal processing (i.e. the hyper beamforming effect) and it is discussed in sub section 5.5. in all cases, the capability of the parameters in reducing SLL and increasing directivity are evaluated. The evaluation of the parameters also considers the system cost in terms of antenna size, antenna weight and system complexity in addition to the radiation pattern flexibility. The tool used for the simulation is Matlab 2013 on intel core i-3, 4GB RAM Window 7 laptop.

### 5.1. Effect of vertical spacing on SLL and directivity

One of the five determining factor of an antenna array is the inter-element spacing of the array elements. As the elliptical rings of ECAA is constructed one above the other in a uniform vertical spacing, different values of random vertical spacing are iteratively searched for the minimum possible SLL reduction.

The dv is determined by the electrical size of the antenna. If the vertical spacing (dv) is very small, the interferences among the elements will be increased which in return decreases the amount of SLL reduction. On the other hand, if the dv very big then the size of the antenna will increase in addition to decreasing the overall radiation intensity of the antenna. The vertical spacing can be searched using successive simulations using try and error while tuning the vertical spacing and observing the SLL reduction. For easier vertical spacing (dv) determination, we have used the following iterative algorithm. When a better SLL (Sn) is achieved by changing the vertical spacing the new dv (dvn) is used with the corresponding new SLL. This will continue until the iteration size (I) becomes zero. Initially we have started the algorithm at the SLL (So), so that we can update it when better reduction is achieved at another new value of the vertical spacing.

Iterative algorithm:

1. Initialize
    a. So, dv, I
2. While I not 0, do {
    a. Calculate SLL
    b. If So>Sn {So = Sn; dv = dvn}
    c. Else {I-1}
    d. End else
    e. End if
    f. End while}
3. Return dv

Finally, after 500 iterations the minimum SLL is converged to be -13.88 dB at dv = 0.9999λ. This side lobe reduction is better than the side lobe resulted at dv = 0.5λ (-8.5 dB) but the FNBW is increased from 38° to 50°. In other words, the beam width is widened from 38° to 50°. This indicates that the directivity is decreased significantly which is not desired characteristic of the radiation pattern. The simulation result shown in Fig. 2 indicates that the SLL and the FNBW are contrasting to each other.

Therefore, in this study, we prefer to proceed with the 0.5λ vertical spacing to take the opportunity of narrow beam width at the expense of small SLL.

### 5.2. Effect of the number of ellliptical rings on SLL & directivity

Fig. 3 indicates that at the expense of the size of the antenna the SLL reduction can be increase, however, the amount of SLL reduction is small. For example, when the size of the antenna increases 3 times (from 2 rings to 6 rings), the amount of SLL reduction is only 3 dB. On the other hand, increasing the size of the antenna is not a desire of communication engineers. Typically, Fig. 1(a) is a 3 ring each having 12 elements.

### 5.3. Effect of eccentricity on SLL and directivity

The normalized AF in dB versus angles in degree is simulated at the same eccentricity of 0.5 and at the same vertical spacing of 0.5λ. The effect of changing the major axis (a) on SLL and Directivity is observed. The major axis is varied from 1.15 to 0.6 and the effect of this major axis variation on the SLL and Directivity is evaluated as shown in Fig. 4.

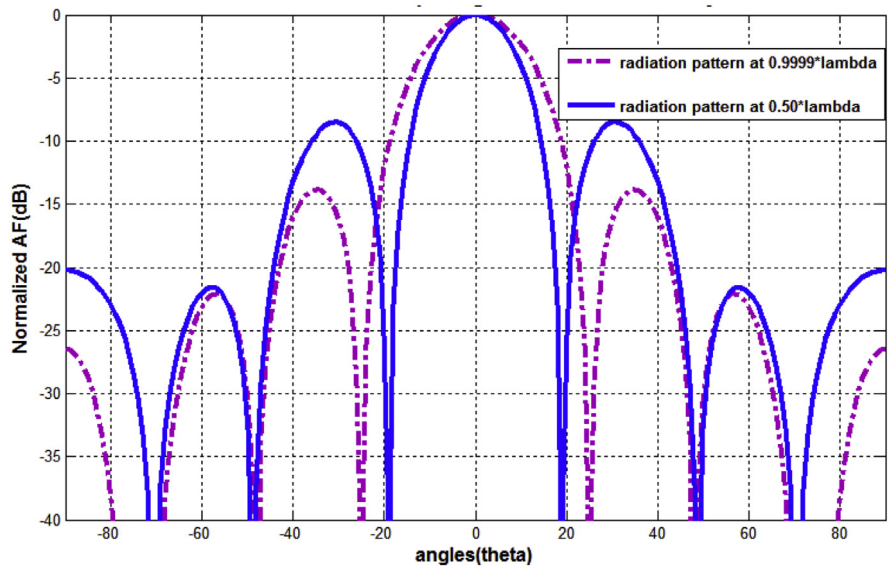

**Fig. 2.** Effect of vertical spacing on SLL and directivity.

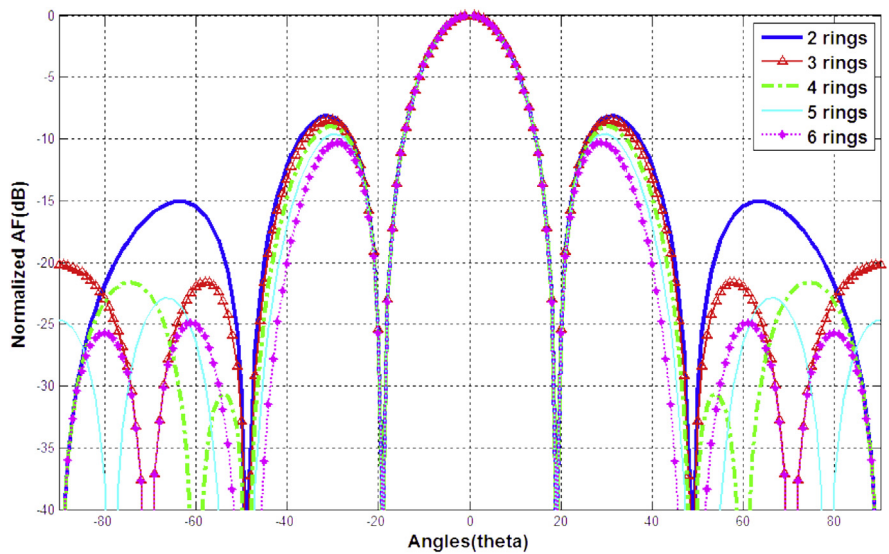

**Fig. 3.** SLL and directivity against the number of elliptical rings each having 12 elements.

Note that varying the ellipse major axis (a) while keeping the minor axis (b) means we are varying the eccentricity of the EcAA. The relationship between major axis (a), minor axis (b) and the eccentricity (e) is given by $b = a\sqrt{1 - e^2}$. At 'a' = 1.15, the SLL is -8.5dB and the BWFN is 38°. When the major axis 'a' decreased from 1.15 to 0.6 (i.e. when the eccentricity is increased from 0.5 to 1.3), the SLL gradually decreased from -8.5dB to -17.22dB and the FNBW widened from 38° to 78°. This result demonstrates that the eccentricity variation results in a SLL reduction of 8.72 dB at the expense of 40° beamwidth increase. The increase in beamwidth implies decrease in the directivity and the amount of SLL reduction is insignificant for critical applications. The result of [14] also confirmed that a net of 3.83dB SLL reduction for eccentricity variation from 0.5 to 0.7. Both this paper's and [14]'s results indicates that significant SLL reduction is not achieved by varying the ellipse eccentricity.

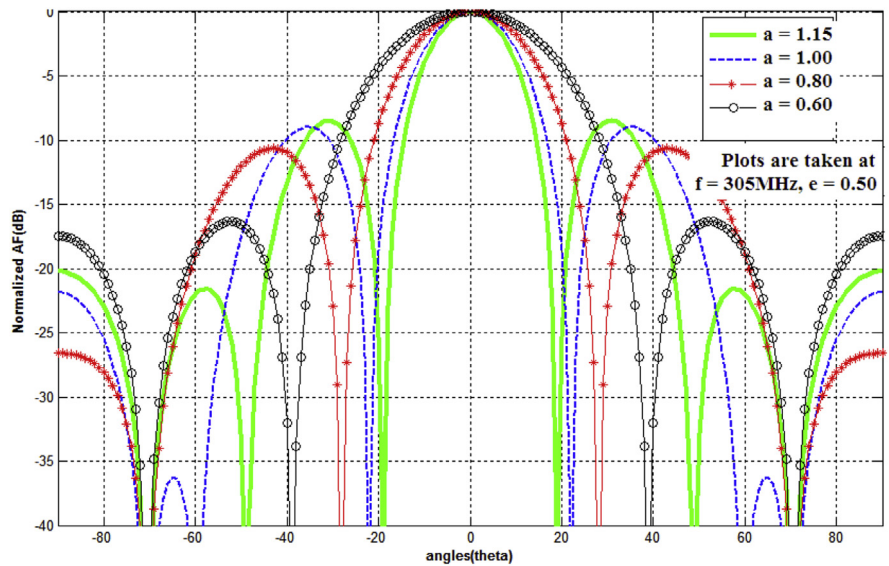

**Fig. 4.** SLL and directivity versus ellipse shape.

### 5.4. Effect of number of elements on SLL and directivity

The effect of the number of elements of the ECAA on the SLL and directivity is studied by keeping the major axis on 1.15 at 305 MHz operating frequency. From the results of Fig. 5, we can observe that the SLL continuously decreases for each doubling the number of array elements. From Fig. 5, good SLL reduction without disturbing the directivity is effectively achieved.

This is very good in preserving the BWFN constant while reducing the SLL unlike the results discussed in Figs. 2, 3, 4. The main difference of sub sections 5.1, 5.2 and 5.3 is: In sub section 5 and 5.2, the level of side lobe reduction is very small and the side lobe reduction disturbs the FNBW (i.e. the directivity of the radiation pattern is decreased when the SLL reduction increases). Whereas in sub section 5.3, significant SLL reduction is achieved without affecting the directivity of the antenna. In sub section 5.3, however, increasing the number of elements is known to increase the array complexity, system cost, power and weight of the antenna. For this reason, in this study, the numbers of antenna arrays are limited to 36 elements. The effect of number of array elements increase on SLL and FNBW is summarized in Table 2.

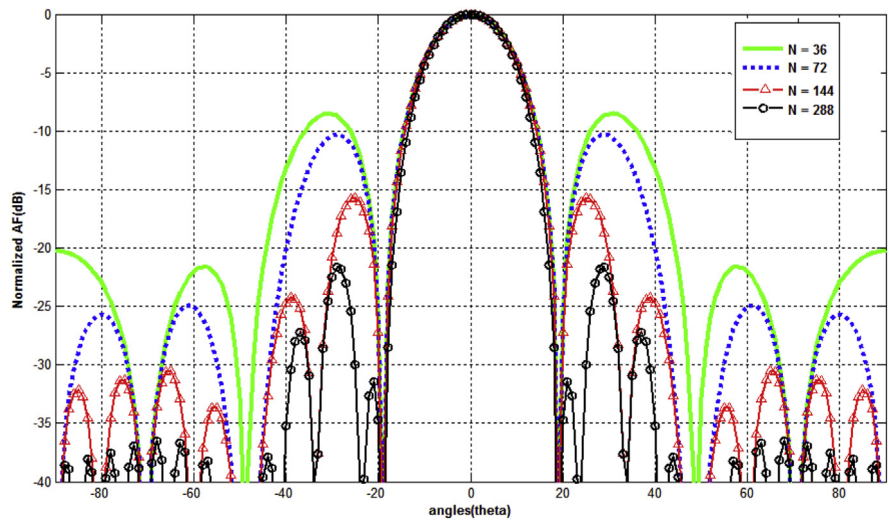

**Fig. 5.** SLL and directivity against the antenna array number of elements.

**Table 2.** Effect of number of elements on SLL and Directivity.

| Experiment number | Number of elements (N) | FNBW (degree) | SLL (dB) |
| --- | --- | --- | --- |
| 1 | 18 | 38 | -06.86 |
| 2 | 36 | 38 | -08.50 |
| 3 | 72 | 38 | -10.30 |
| 4 | 144 | 38 | -15.72 |
| 5 | 288 | 38 | -21.66 |

### 5.5. Effect of hyper beam on ECAA performance

In order to evaluate the effect of hyper beam on ECAA performance, the influence of hyper beam exponent on SLL reduction and on increasing the directivity is explored. A normalized Array Factor (AF) versus the elevation angles are simulated for different values of the hyper beam exponent. The effect of uniform excitation current on SLL and Directivity at different values of the hyper beam exponent is simulated in Fig. 6.

The rectangular plot, shown in Fig. 6, is for uniform excitation current having the same amplitude with 36 numbers of elements. As the hyper beam exponent (r) decreases from 0.3 to 0.1, the SLL reduced from -17.07dB to -41.76 dB without affecting the BWFN. When the -41.76 dB SLL reduction is expressed in percentage, a 99.18% of SLL reduction over the isotropic antenna and 62.416% of SLL reduction over the same antenna array without hyper beam is accomplished. This is one of the major contributions of this work as the SLL is decreased much below the peak value of the main beam (i.e., below the isotropic antenna).

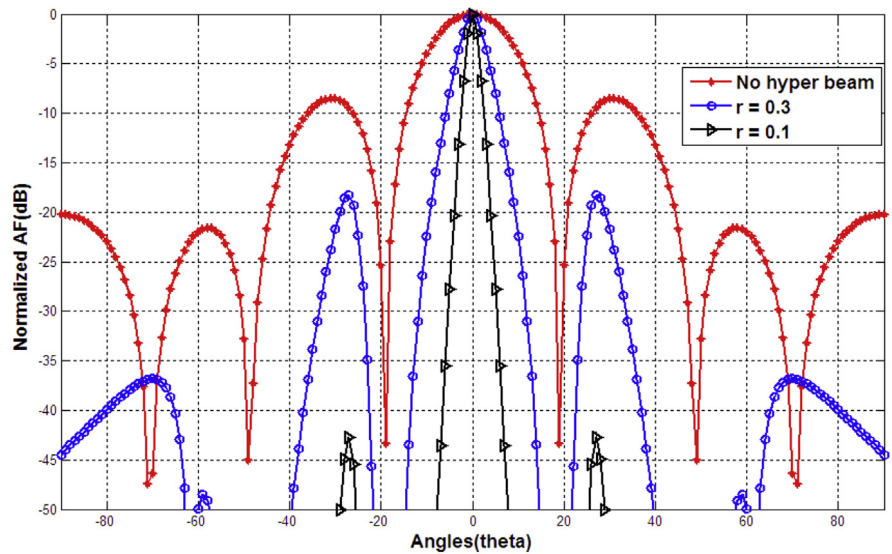

**Fig. 6.** SLL and Directivity variation against hyper beam exponent, r.

The hyper beam exponent is a spatial signal processing parameter that varies between 1 and 0.1. If we continue varying the hyper beam exponent below 0.1, it results in further SLL reduction. However, as the hyper beam exponent below 0.1 results in grating lobe, it is not recommended. Fig. 6 indicates that the SLL reduction is achieved by varying the hyper beam exponent. Specifically, the SLL has shown significant reduction gradually when the hyper beam exponent decreases.

Unlike sub sections 5.1, 5.2, 5.3 and 5.4, the hyper beam exponent method of antenna design is achieved with the minimum possible system cost. More briefly, the SLL reduction is achieved neither with increasing the number of antenna array elements nor with disturbing the directivity of the radiation pattern. The SLL reduction and directivity increment are achieved simultaneously. As the hyper beam exponent processes the beams of the antenna array, it doesn't affect the weight, size and shape of the antenna. This concludes the hyper beam exponent technique of antenna pattern synthesis is ideal for practical applications. In general, the flexibility of the radiation pattern can be controlled by changing the hyper beam exponent. This is shown in Table 3.

**Table 3.** The effect of hyper beam on SLL and directivity.

| Exp No. | SLL and Directivity variation technique | FNBW | SLL (dB) | HPBW |
|---|---|---|---|---|
| 1 | Ordinary radiation pattern | 38° | -8.50 | 17.8° |
| 2 | Hyper beam, r = 1.0 | 38° | -9.07 | 13.4° |
| 3 | Hyper beam, r = 0.5 | 38° | -12.54 | 8.1° |
| 4 | Hyper beam, r = 0.3 | 38° | -17.07 | 4.5° |
| 5 | Hyper beam, r = 0.1 | 38° | -41.76 | 2.6° |

From Fig. 7, the polar plots are observed to cover from -90° to 90° in the elevation with full beam scan of 60° (from -330° to 30°) and the azimuthal angle covers the whole 360°. This indicates the proposed antenna array is really satisfactory for modern radars, deep space exploration and satellite communications, providing scanning over 360°.

A summary of the advantage and disadvantages of the different parameters discussed in Sub-sections (Ss) 5.1, 5.2, 5.3, 5.4 and 5.5 are given in Table 4. To sum up, paper [14] has worked on ECAA to design the antenna with minimum SLL with the capabilities to scan the whole hemisphere and it has successfully reduced the SLL using PSO. However, this SLL reduction is insufficient to be effectively utilized in high signal quality seeking applications like in radar systems, satellite communication and deep space explorations. Furthermore [19], has used symmetric weighted thinned array with pattern reconfigurable antenna as elements which results in good degrees of freedom in array design that reduces the SLL. However, only 11.1 dB of SLL is achieved which is insufficient for long distance communication applications. The SLL and FNBW resulted in this paper are better than the papers [10] through [19]. The hyper beamforming-based SLL reduction, FNBW reduction and radiation pattern flexibility results are significant compared to the different techniques mentioned in the literatures including GA, PSO and SaDE. However, it is possible to enhance the antenna design further if the hyper beamforming is hybridized with other optimization algorithms. The significance of hyper beamforming for radiation pattern flexibility is justified as the hyper beamforming have one more parameter called hyper beam exponent for adjusting the radiation pattern.

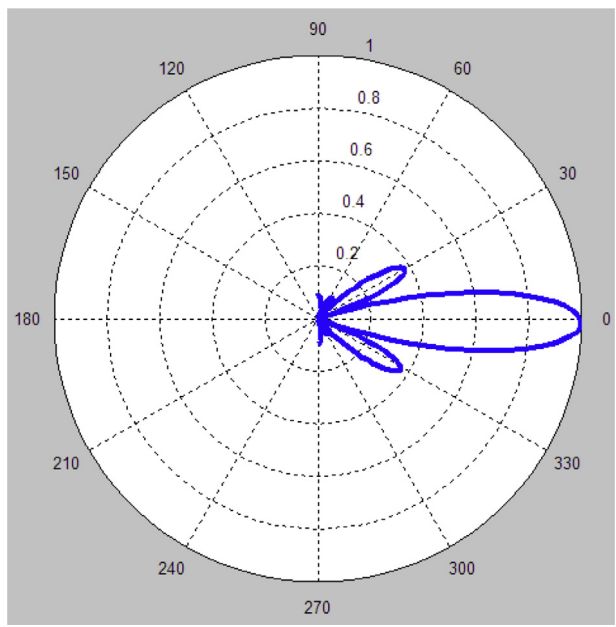

**Fig. 7.** Polar plot of scanning with hyper beam.

Heliyon

Article No~e01230Table 4. Summary of pattern synthesis parameters.

| Ss | Parameter | Advantage | Disadvantage |
|---|---|---|---|
| 5.1 | Vertical element spacing (rings separation) | ■ Some SLL reduction<br>■ Some pattern flexibility<br>■ Simple to design | ■ Decrease directivity<br>■ Static design (less flexible because it is physical component of the antenna) |
| 5.2 | Number of elliptical rings | ■ Some SLL reduction | ■ The size of the antenna increases<br>■ As the total number of elements increases the weight, power and complexity increases |
| 5.3 | Eccentricity of the ellipse | ■ SLL reduction is possible<br>■ Pattern flexibility is possible | ■ Decrease directivity<br>■ Static design (less flexible because it is physical component of the antenna) |
| 5.4 | Increase the number of array elements | ■ Very good SLL reduction<br>■ Doesn't decrease or increase directivity<br>■ Little pattern flexibility | ■ Antenna size<br>■ Antenna weight<br>■ System complexity<br>■ Increase power<br>■ Less flexible because it affects number of elements & hence the physical size of the antenna |
| 5.5 | Hyper beam exponent | ■ Excellent SLL reduction<br>■ Better directivity<br>■ Excellent radiation pattern flexibility<br>■ Enables electronic beam steering<br>■ Doesn't affect the physical configuration of the antenna | ■ Moderate design complexity (because of processing power of the beamforming circuit) |

## 6. Conclusion

The SLL and directivity of the proposed antenna array is evaluated using the geometrical parameters-ellipse eccentricity, element-spacing and number of elements. From these results, simultaneous increase in directivity and SLL reduction is not achieved without increasing the number of elements. This is not an ideal antenna pattern synthesis requirement of wideband applications because it makes the system complex and bulky. On the other hand, the performance of ECAA is investigated using hyper beamforming and the simulation has shown that an excellent SLL reduction along with good directivity. Furthermore, radiation pattern flexibility is successfully achieved by the hyper beam exponent variation. Besides, an elevation and azimuth planes scanning are possible to cover wide range of coverage that is a blueprint for wideband applications.

14  https://doi.org/10.1016/j.heliyon.2019.e012302405-8440/© 2019 The Authors. Published by Elsevier Ltd. This is an open access article under the CC BY-NC-ND license
(http://creativecommons.org/licenses/by-nc-nd/4.0/).

## Declarations

### Author contribution statement

Gebrehiwet G. Lema: Conceived and designed the experiments; Performed the experiments; Wrote the paper.

Dawit H. Hailu: Analyzed and interpreted the data; Contributed reagents, materials, analysis tools or data.

### Funding statement

This research did not receive any specific grant from funding agencies in the public, commercial, or not-for-profit sectors.

### Competing interest statement

The authors declare no conflict of interest.

### Additional information

No additional information is available for this paper.